\documentclass[aps,prl,showpacs,groupedaddress]{revtex4}

\usepackage[dvips]{graphicx}
\usepackage{dcolumn}

\begin{document}


\title
{Vortex lattice structure dependent on pairing symmetry in Rashba 
superconductors
}

\author{Norihito Hiasa, Taro Saiki, and Ryusuke Ikeda}

\affiliation{%
Department of Physics, Kyoto University, Kyoto 606-8502, Japan
}

\date{\today}

\begin{abstract} 
Vortex lattice structures in Rashba noncentrosymmetric superconductors in magnetic fields parallel to the basal plane (${\bf H} \perp c$) are examined based on the BCS-like Hamiltonian and the resulting Ginzburg-Landau functional. Due to the momentum dependent anisotropy of the Zeeman effect induced by the broken inversion symmetry, the vortex lattice in higher fields generally shows some unidirectional modulation of Fulde-Ferrell-Larkin-Ovchinnikov (FFLO) type orienting in the plane perpendicular to ${\bf H}$. However, the direction of the modulation and the lattice structure depend significantly on the underlying pairing symmetry: When the mixing between spin singlet and triplet pairing components is negligible, the resulting modulated structure tends to have reflection symmetry, while the vortex lattice in systems with a significant singlet-triplet mixing has no reflection symmetry in most cases. The latter result implying the presence in {\it real} materials of two degenerate orientations of the lattice structure separated by domain walls may be relevant to the extremely low magnetic decay rate observed in CePt$_3$Si. 
\end{abstract}



\maketitle

\section{I. Introduction} 

The Pauli paramagnetism has crucial effects on superconducting vortex states. Reflecting the pictures \cite{MT,FF,LO} expected in the vortex free (Pauli) limit, it tends to change the character of the {\it mean field} superconducting transition occurring on the depairing field $H_{c2}(T)$ from the conventional second order into a first order one \cite{AdachiIkeda} and induces an additional spatial modulation \cite{AdachiIkeda,IkedaNawata} as a reflection of a Fulde-Ferrell-Larkin-Ovchinnikov (FFLO) state in the Pauli limit. Further, such a spatial variation of the pair-field, i.e., superconducting order parameter, due to the vortices and the FFLO modulation can also induce a parity mixing \cite{Nagai,Lebed} in the vortex state. In the ordinary superconductors with inversion symmetry, however, these effects are quite weak, and, in particular, it is not expected that the field-induced parity mixing \cite{Nagai,Lebed} changes the vortex phase diagram qualitatively. Besides this, effects of the paramagnetic depairing on the vortex lattice structure have not been discussed until recently. It has been found \cite{HiasaIkeda} that, in systems where the paramagnetic effect is strong enough, it weakens an anisotropy in the vortex lattice structure reflecting the pairing symmetry or the band structure and stabilizes the isotropic triangular lattice structure.

In this paper, we study possible relations between the pairing symmetry and the vortex lattice structure in {\it noncentrosymmetric} superconductors with spin-orbit coupling of Rashba type in magnetic fields parallel to the basal plane (${\bf H} \perp c$). It is found that, in contrast to the above-mentioned consequences in centrosymmetric superconductors with inversion symmetry, spatial modulations induced by the paramagnetic depairing and a field-induced parity mixing occur even for relatively lower values of the Maki parameter \cite{Maki} $\sqrt{2} H_{\rm orb}(0)/H_P(0)$, and that the vortex lattice structure in Rashba superconductors strongly depends upon the details of the orbital component of the Cooper pairing state, where $H_{\rm orb}(0)$ and $H_P(0)$ are the orbital and Pauli limiting fields at zero temperature, respectively. The main origin of these intriguing effects is an anisotropic Zeeman energy for quasiparticles stemming from the antisymmetric spin-orbit coupling (ASOC) peculiar to noncentrosymmetric superconductors. This ASOC is expressed as an additional term  
\begin{equation}\label{eq:hsoc}
{\cal H}_{\rm soc}= \sum_{{\bf k}, \alpha, \beta} c_{{\bf k}, \alpha}^\dagger \, \zeta \, {\bf g}_{\bf k}\cdot{\bf \sigma}_{\alpha, \beta} \, c_{{\bf k}, \beta}
\end{equation}
in the electronic Hamiltonian, where $\zeta$ is the energy scale measuring the magnitude of ASOC, and $c_{{\bf k},\alpha}$ is the annihilation operator of conduction electron with momentum ${\bf k}$ and spin projection $\alpha$. The type of ASOC is defined by the vector ${\bf g}_{\bf k}$ corresponding to the Fourier transform of the ASOC, and, in Rashba superconductors, it can be expressed as $({\bf k} \times {\hat z})/k_F$, where $k_F$ is the Fermi wavenumber, and the $c$-axis corresponds to the $z$-direction. Due to this term, the quasiparticle band of our interest in the normal state is split into two pieces \cite{GR}. The strength of ASOC is measured by the dimensionless quantity 
\begin{equation}\label{eq:deltan}
\delta N \equiv \frac{N_2 - N_1}{N_1+N_2} \sim \frac{|\zeta|}{E_F},
\end{equation}
expressing a normalized difference between the density of states on the split two bands, $N_1$ and $N_2$. Hereafter, each Fermi surface (FS) or band will be specified by the indices $a=1$ and $2$. Through the present paper, the ideal limit $\delta N \to 0$ corresponding to the limit of large band width ($E_F \to \infty$) is often considered in order to understand physical origins of structural changes of the vortex lattice. As a result of the coupling in the spin space between the ASOC and the original Zeeman terms, the quasiparticle Zeeman energy on the split FSs becomes ${\bf k}$-dependent. Further, we focus in this article on 
noncentrosymmetric superconductors satisfying the condition \cite{FAKS} 
\begin{equation} \label{eq:conASOC}
  {\rm max}(T,\mu H) \ll \zeta \ll E_F,  
\end{equation}
where $\mu H$ is the Zeeman energy in centrosymmetric case. Under this condition, the Zeeman energy in noncentrosymmetric materials becomes highly anisotropic in the momentum space so that the paramagnetic depairing is ineffective in ${\bf H} \parallel c$ \cite{FAKS}, where other high order corrections in $\mu H/\zeta$ were neglected. 
This ASOC-induced anisotropy in the Zeeman energy makes emergence of modulated vortex states in Rashba superconductors easier compared with that in centrosymmetric systems. 
\begin{figure}[t]
\scalebox{1.8}[1.8]{\includegraphics{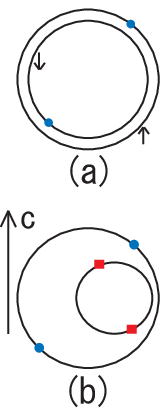}}
\caption{(Color online) Fermi surfaces (FSs) in a magnetic field of (a) a centrosymmetric superconductor and (b) a Rashba noncentrosymmetric one. The colored dots indicate partners of Coooer pairings on the FSs. In the case (b) satisfying eq.(\ref{eq:conASOC}), each Cooper pair is formed on the same FS, and the paramagnetic depairing is effective only through the {\it relative} displacement of the two FSs of which the magnitude is $2 Q_0 = 2 \mu H/v_F$ (see sec.I). In the limit in which one of the two FSs is irrelevant to superconductivity, such an FS-displacement induced by the Zeeman effect is trivial and reduces to the orbital-limited case.} \label{fig.1}
\end{figure}

Before proceeding further, we introduce two key parameters for determining the type of modulations in vortex states in Rashba superconductors. One is $\delta N$ defined in eq.(\ref{eq:deltan}). However, we note that, as indicated in Fig.1(b), a modulation induced by the paramagnetic depairing is expected even in $|\delta N| \to 0$ limit. The other is a parameter measuring closeness between the most attractive singlet and triplet pairing states and will be expressed in the 
form 
\begin{equation}\label{eq:deltaw}
  \delta w = \frac{2(N_1 +N_2)}{(w^{-1})_{tt}-(w^{-1})_{ss}},
\end{equation}
where $w_{ss}$ ($w_{tt}$) is the strength of attractive interaction in the most attractive pairing state in singlet (triplet) channels. In the limiting case with just a single pairing state, $|\delta w|$ vanishes, while it is divergent when both of the two pairing channels equally contribute to superconductivity. In the latter case, only one of two FSs contributes to the pairing. Then, the pure orbital-limiting case is reached, because a field-induced displacement of {\it the only} FS participating in superconductivity does {\it not} affect the pairing itself (see Fig.1(b)). 

\begin{figure}[t]
\caption{(Color online) Typical $h$-$t$ phase diagrams of (a) a centrosymmetric superconductor and (b) a Rashba noncentrosymmetric one in $\delta N \to 0$ limit following from the present theory, where $h$ and $t$ are normalized field and temperature, respectively. Each blue curve denotes the $H_{c2}(T)$-curve, while each red curve is a first order structural transition (FOST) between different structures of the vortex lattice. In (a), the possibility of a discontinuous $H_{c2}$-transition and an FFLO modulation parallel to the field was neglected to make comparison between (a) and (b) easier. Insets: The images (A), (B), and (C) in each figure are real space patterns of the amplitude of the pair-field in the $x$-$z$ plane at each ($h$, $t$) below $H_{c2}$-curve. The energy gap nearly vanishes in the darkest regions. } \label{fig.2}
\end{figure}

To clarify why a state modulating in the plane perpendicular to ${\bf H}$ tends to occur {\it more easily} compared with in the centrosymmetric case, the $h$-$t$ phase diagrams (a) in the {\it centrosymmetric} case and (b) in Rashba case in $\delta N \to 0$ limit are compared with each other in Fig.2, where $t=T/T_c$, and the normalized field $h$ will be defined in the ensuing sections. For simplicity, the possibility \cite{AdachiIkeda} of the first order $H_{c2}$-transition and of a modulation {\it parallel} to ${\bf H}$ in Fig.2(a) is neglected here. Between the figures (a) and (b), the only difference in the used Hamiltonian is the Zeeman energy term: In (a), the ordinary isotropic Zeeman energy $\mu {\bf H}\cdot{\bf \sigma}$ is used, where ${\bf \sigma}$ denotes the spin projection, and the modulated state, (A) and (B), corresponding to the FFLO state in the Pauli limit appears in higher fields and is separated by a first order transition occurring on the red curve from the conventional triangular lattice (C). On the other hand, in (b), the Zeeman energy is influenced by ${\cal H}_{\rm soc}$ and takes the form $\mu {\bf H}\cdot({\hat {\bf k}} \times {\hat z})$. Due to this anisotropy in the momentum space of the Zeeman effect, the overall effect of the paramagnetic depairing is weakened in the noncentrosymmetric case, as can be seen from the relative enhancement of $H_{c2}$ and the narrower FFLO region with the structure (A) in (b). However, novel modulated vortex states, the intermediate states (B) and (C), tend to occur in (b) due to the momentum dependent Zeeman energy term. That is, contrary to the centrosymmetric case with isotropic Zeeman energy, a novel modulated vortex state can be expected to occur in Rashba superconductors even if the paramagnetic depairing effect is so weak that the conventional FFLO-like modulated state (A) does not appear. In fact, this is the basic reason why, as will be shown below, various spatially-modulated states can occur dependent on the pairing symmetry in Rashba superconductors. 

It should be stressed that the novel spatially-modulated states, (B) and (C) in (b) in intermediated fields, appear irrespective of $\delta N$ and hence, is not relevant to the helical {\it phase} modulation \cite{KAS,Samo} argued to occur in the vortex {\it free} situation in Rashba superconductors. The phase modulation is hidden in the vortex states in Rashba case and merely appears as an anisotropy orienting the direction of the novel modulated structure. However, the helical phase modulation is observable in noncentrosymmetric superconductors of cubic type and should be detected as a local and {\it transverse} magnetization \cite{HI}. 

In sec.II, the microscopic basis of our calculation is explained. After deriving the Ginzburg-Landau (GL) free energy functional, results on phase diagrams in the pure singlet or pure triplet case are shown and discussed separately for a couple of pairing states in sec.III. In the next section, the analysis is extended to the case with a singlet and triplet mixing, and discussions on relevance of the present results to real systems are given in sec.V. 

\section{II. Electronic Hamiltonian}

We start from the following electronic Hamiltonian 
\begin{equation} \label{eq:NCSH}
  {\cal H}_{\rm el}={\cal H}_{\rm single}+{\cal H}_{\rm int},
\end{equation}
where 
\begin{equation} \label{eq:defNCSHsingle}
  {\cal H}_{\rm single}=\sum_{{\bf k},s_1,s_2} c^{\dagger}_{{\bf k},s_1} \left( \varepsilon _{\bf k} \delta _{s_1 ,s_2}
  +(\zeta ({\bf g}_\nu)_{\bf k} +\mu _B H_\nu) (\sigma_\nu)_{s_1 ,s_2} \right) c_{{\bf k},s_2} 
\end{equation}
is the kinetic energy Hamiltonian accompanied by ${\cal H}_{\rm soc}$ and the bare Zeeman term, $c^{\dagger}_{{\bf k},s}$ is the creation operator of a conduction electron with the wave vector ${\bf k}$ and the spin index $s$, 
$\mu H$ the magnitude of the bare Zeeman energy, $H=|{\bf H}|$, $\varepsilon_{\bf k}$ is the bare dispersion of conduction electrons defined with no ASOC, 
$(\sigma_\mu)_{s_1,s_2}$ is a Pauli matrix, and ${\bf g}_{\bf k} ={\bf k}\times \hat{z}/k_F$ (see sec.I). At the present stage, the orbital effect of the magnetic field was neglected in eq.(\ref{eq:defNCSHsingle}). 

Throughout this paper, our calculation is performed within models based on the quasi two-dimensional (Q2D) dispersion of quasiparticle energy
\begin{equation} \label{eq:RFSenergy}
\varepsilon_{\bf k} =\frac{1}{2m}(k_x^2 +k_y^2) +J(1-{\rm cos}({\tilde k}_z)),
\end{equation}
where ${\tilde k}_z \equiv k_z d$, and $d$ is the period in the $z$ (or $c$) direction. In most of real tetragonal noncentrosymmetric materials, the main Fermi surface (FS) seems to be an ellipsoid. However, effects of discrete layer structure in the original Q2D model are not considered hereafter by assuming $d \ll \xi_z(0)$ so that $d$ does not appear in the resulting GL functional, and, under this simplification, the GL functional for the Q2D FS is equivalent to that of the ellipsoidal FS, where $\xi_\nu(0) \ (\nu=x,y,z)$ is the zero temperature coherence length in the $\nu$-direction. Then, the only parameter characterizing FS effects is the anisotropy parameter $\gamma$ between the coherence lengths, which is given in terms of ${\tilde J} \equiv J/\varepsilon_F$ by 

\begin{equation} \label{eq:FSaniso}
  \gamma =\frac{\xi_x(0)}{\xi_z(0)} =\sqrt[]{\mathstrut \frac{\langle v_x^2 \rangle}{\langle v_z^2 \rangle}}
  =\frac{2\ \sqrt[]{\mathstrut 1-\tilde{J}}}{\pi\tilde{J}}.
\end{equation}

For this FS, the average of a quantity $f({\bf v})$ over the momentum {\it on} the FS is given by 
\begin{eqnarray}
  && \Bigl\langle f({\bf v}) \Bigr\rangle =\int_{-\pi}^{\pi} \frac{d{\tilde k}_z}{2\pi} \int_0^{2\pi} \frac{d\phi_{\bf k}}{2\pi} f({\bf v})
 \\
  && \quad v_x =v_F\ \sqrt[]{\mathstrut 1-\tilde{J}(1-{\rm cos}{\tilde k}_z)}{\rm cos}\phi_{\bf k}
 \nonumber \\
  && \quad v_y =v_F\ \sqrt[]{\mathstrut 1-\tilde{J}(1-{\rm cos}{\tilde k}_z)}{\rm sin}\phi_{\bf k}
 \nonumber \\
  && \quad v_z =J \, d \, {\rm sin}{\tilde k}_z,
 \nonumber
\end{eqnarray}
where $\phi_{\bf k}= {\rm tan}^{-1}(k_y/k_x)$. Hereafter, we assume $k_F d = \pi$ in order to merely reduce the number of inessential material parameters. 

The interaction Hamiltonian takes the following generic form
\begin{equation} \label{eq:NCSHint}
  {\cal H}_{\rm int}=\frac{1}{V}\sum^{}_{\bf p,k_1,k_2} W_{\alpha \beta ,\gamma \delta} ({\bf k_1,k_2}) \, c^{\dagger}_{{\bf k_1+p}/2,\alpha} \, c^{\dagger}_{-{\bf k_2+p}/2,\beta} \, c_{-{\bf k_2+p}/2,\delta} \, c_{{\bf k_2+p}/2,\gamma},  
\end{equation}
where the interaction potential $W_{\alpha\beta,\gamma\delta}({\bf k}_1,{\bf k}_2)$ may be expressed as 
\begin{equation} \label{eq:NCSint}
  W_{\alpha \beta ,\gamma \delta}({\bf k}_1 ,{\bf k}_2 ) =-\frac{1}{2}\sum^{}_{i,j=s,t}w_{ij}
  \Bigl( {\tau}^\dagger_i({\bf k}_1) \Bigr)_{\alpha \beta} \Bigl( {\tau}_j({\bf k}_2) \Bigr)_{\delta \gamma}
\end{equation}
with $\tau_s({\bf k})={\rm i}\sigma_y {\hat \Delta}_{\bf k}, \ \tau_t({\bf k})={\rm i}(\sigma_y\sigma_\mu)\cdot({\bf g}_{\bf k})_\mu {\hat \Delta}_{\bf k}$, where ${\hat \Delta}_{\bf k}$ expresses a normalized pairing function of the dominant component in the spin-singlet channel, and the so-called $d$-vector of the spin-triplet component was replaced in $\tau_t$ by ${\bf g}_{\bf k}$ based on the inequality (\ref{eq:conASOC}) \cite{FAKS}. Further, ${\hat {\bf o}}$ denotes the unit vector parallel to ${\bf o}$. Hereafter, we will focus on the case in which both of the singlet ($s$) and triplet ($t$) channels are attractive, and the matrix $w_{i,j}$ ($i$, $j=s$ and $t$) is positive definite. 

Due to ASOC, ${\cal H}_{\rm sing}$ is not diagonalized via the spin states, and the following unitary transformation is needed to diagonalize it: 
\begin{eqnarray}
  U_0^\dagger({\bf k}) \left(
   \begin{array}{ccc}
    c_{{\bf k},\uparrow} \\
    c_{{\bf k},\downarrow} \\
   \end{array}
  \right)
  =\left(
   \begin{array}{ccc}
    f_{{\bf k},1} \\
    f_{{\bf k},2} \\
   \end{array}
  \right),
\end{eqnarray}
where $f$ is the field operator of the resulting quasiparticles, and 
\begin{equation}
  U({\bf k}) =\frac{1+ {\rm i}({\rm sin}\phi_{\bf k}\sigma_y -{\rm cos}\phi_{\bf k}\sigma_x)}{\sqrt[]{\mathstrut 2}}. 
\end{equation}
Then, ${\cal H}_{\rm single}$ is represented in terms of quasiparticle states $f_{{\bf k},a}$ with two split FSs 
\begin{equation} \label{eq:NCSHsingle}
  {\cal H}_{\rm single}=\sum_{\bf k}\sum_{a=1,2} f_{{\bf k},a}^\dagger E_{{\bf k},a} f_{{\bf k},a}, 
\end{equation}
where 
\begin{equation}\label{eq:QPenergy}
  E_{{\bf k},a}=\varepsilon_{\bf k} +(-1)^{a+1} |\zeta {\bf g}_{\bf k}+\mu {\bf H}|. 
\end{equation}
The quasiparticle Green's function close to FS $a$ is 
\begin{equation} \label{eq:NCSGreenfunc}
  {\cal G}_a ({\bf k},{\rm i}\epsilon) \simeq \frac{1}{{\rm i}\epsilon -\xi_a + (-1)^a \mu {\bf H}\cdot{\hat {\bf g}}_{\bf k}},
\end{equation}
where $\xi_a$ is the single particle energy measured from the FS $a$, and ${\hat {\bf g}}_{\bf k} = {\bf g}_{\bf k}/|{\bf g}_{\bf k}|$. Throughout this paper, the field configuration ${\bf H} \parallel \hat{y}$ is assumed in which the Pauli paramagnetism is effective, and the ${\bf H} \parallel {\hat z}$ (${\bf H} \parallel c$) configuration in which the Zeeman term vanishes will not be considered. 

Correspondingly, the interaction Hamiltonian is expressed in the form 
\begin{equation} \label{eq:NCSdiaHint}
  {\cal H}_{\rm int}=-\frac{V}{2}\sum^{}_{\bf p}\sum_{i,j=s,t}
  w_{ij} \left( \Psi ^{(i)}_{\bf p} \right)^\dagger \Psi ^{(j)}_{\bf p}, 
\end{equation}
where the pair-field operators take the form
\begin{eqnarray} \label{eq:CSingletField}
  &&\Psi^{(s)}_{\bf p}=-\sum^{}_{\bf k} \frac{1}{V} \sum^{}_{a=1,2} {\hat \Delta}_{\bf k} \, e^{[{\rm i}(-1)^{a+1}\phi_{\bf k}]} \, 
  f_{-{\bf k}+{\bf p}/2,a}  \, f_{{\bf k}+{\bf p}/2,a}
 \\ \label{eq:CTripletField}
  &&\Psi^{(t)}_{\bf p}=\sum^{}_{\bf k}\frac{|{\bf g}_{\bf k}|}{V}\sum^{}_{a=1,2} {\hat \Delta}_{\bf k} \, e^{[{\rm i}(\pi(a+1)-(-1)^a \phi_{\bf k})]} \,
  f_{-{\bf k}+{\bf p}/2,a} \, f_{{\bf k}+{\bf p}/2,a}. 
\end{eqnarray}

\section{III. Single pairing case} 

First, let us start from explaining our results in pure singlet cases where $w_{tt}=w_{st}=w_{ts}=0$. The following results remain essentially valid for the corresponding triplet-only cases. In this section, we consider the cases with (i) a full-gap ${\hat \Delta}_{\bf k}=1$, (ii) horizontal-line gap nodes ${\hat \Delta}_{\bf k} = \sqrt{2} {\rm cos}(2{\tilde k}_z)$, and (iii) vertical-line gap nodes ${\hat \Delta}_{\bf k}= \sqrt{2} {\rm cos}(2\phi_{\bf k})$. The cases (i) and (iii) correspond to the ordinary $s$-wave pairing and $d_{x^2-y^2}$-pairing ones, respectively, while a pairing state \cite{Tada} proposed for CeRhSi$_3$ \cite{Kimura} and CeIrSi$_3$ \cite{Settai} corresponds to the case (ii). 

According to the familiar Hubbard-Stratonovich transformation \cite{HS}, the quadratic term of the GL functional is given by 
\begin{eqnarray} \label{eq:singletfree}
F_2^{(s)} &=& \int d^3r \biggl[ \, \frac{1}{w_{ss}} |\Delta_s|^2 - \sum_{a=1,2} \Delta_s^* K^{(a)}_2({\bf \Pi}) \Delta_s \, \biggr] \nonumber \\ 
&=& N \int d^3r \Delta_s^* \biggl[ \, \frac{1}{N w_{ss}} - \int_{\rho_c} d\rho \, \frac{f(\rho)}{2} \sum_{\sigma= \pm 1} \langle |{\hat \Delta}_{\bf k}|^2 ({\rm cos}(\rho v_x Q_0) - {\rm i} \delta N {\rm sin}(\rho v_x Q_0 \sigma)) \exp(-{\rm i}\rho \sigma {\bf v}\cdot{\bf \Pi}) \rangle \biggr] \Delta_s, 
\end{eqnarray}
where 
\begin{equation}
f(\rho) = \frac{2 \pi T}{{\rm sinh}(2 \pi T \rho)},
\end{equation}
$N=N_1+N_2$, 
and $Q_0=2 \mu H/v_F$ is the familiar modulation wavenumber of the {\it vortex-free} FFLO state in centrosymmetric case in low $T$ limit \cite{LO,FF}. In writing eq.(20), the quantity $Q_0/\sqrt{1-\tilde{J}(1-{\rm cos}{\tilde k}_z)}$ was simply expressed as $Q_0$ by assuming a nearly cylindrical FS with a small ${\tilde J}$ ($< 1$). Validity of this treatment will be explained in relation to Fig.3 and also in sec.IV. The kernel $K^{(a)}_2$ is expressed in terms of ${\cal G}_a({\bf k}, {\rm i}\varepsilon)$ by 
\begin{eqnarray} \label{eq:defRKs}
  K_2^{(a)}({\bf \Pi}) &=& T \sum_{\varepsilon} \int_{\bf k} |{\hat \Delta}_{\bf k}|^2 {\cal G}_a({\bf k}, {\rm i}\varepsilon) {\cal G}_a(-{\bf k}+{\bf \Pi}, -{\rm i}\varepsilon) \nonumber \\ 
&=& \frac{N_a}{2} \int_{\rho_c}^{\infty} d\rho f(\rho) \sum_{\sigma=\pm 1} \langle |{\hat \Delta}_{\bf k}|^2 
 \exp(-{\rm i} \sigma \rho {\bf v}\cdot({\bf \Pi}+ (-1)^{a}Q_0 {\hat x})) \rangle, 
\end{eqnarray}
where ${\bf \Pi} = -{\rm i}\nabla + 2 e {\bf A}$ is the gauge-invariant operator for Cooper-pairs, and $\xi_a$ is the normal quasiparticle energy measured from the FS $a$ in zero field. 

Due to the gauge-coupling through ${\bf \Pi}$, the pair-field $\Delta_s({\bf r})$ will be expanded via basis functions of Landau levels (LLs). Before processing the above expression further, the corresponding LL basis function $\varphi_n(z,x|{\bf 0})$ in the gauge ${\bf A} = H z {\hat x}$ will be first determined following the conventional manner of incorporating the anisotropy $\gamma$ in the low field GL region \cite{RIphysicac}: The exponential operator in eq.(22) is rewritten in terms of the identity $e^{A+B} =e^{[A,B]/2} e^A e^B$ with a constant $[A,B]$ as 
\begin{eqnarray}\label{eq:fomROperator}
  \exp(-{\rm i}\rho {\bf v}\cdot{\bf \Pi}) &=& \exp(\rho \mu \Pi_{+} -\rho \mu^\ast \Pi_{-}) 
 \nonumber \\ 
  &=& \exp(-|\mu|^2 \rho^2/2) \, \exp(\rho \mu \Pi_{+}) \, \exp(-\rho \mu^\ast \Pi_{-}), 
\end{eqnarray}
where 
\begin{equation}
  \mu \equiv \frac{1}{\sqrt[]{\mathstrut 2}r_H} (\gamma^{\frac{1}{2}}v_z -{\rm i}\gamma^{-\frac{1}{2}}v_x),
\end{equation}
and 
\begin{equation}
  \Pi_{\pm} =\frac{r_H}{\sqrt[]{\mathstrut 2}} (\gamma^{\frac{1}{2}} \Pi_{x} \mp {\rm i}\gamma^{-\frac{1}{2}} \Pi_{z} )
\end{equation}
are the creation and annihilation operators of LLs acting on the LL basis functions $\varphi_n (z, x|{\bf 0})$, where $r_H=1/\sqrt{2e H}$ is the averaged vortex spacing originating from the flux quantization. 
Then, the basis functions take the form\begin{eqnarray} \label{eq:RLLfunc}
  \varphi_n(z,x|{\bf 0}) &=& N_0 \frac{(-1)^n}{\sqrt[]{\mathstrut 2^n n!}} \sum_{m} H_n
  \left( \frac{\gamma^{\frac{1}{2}}}{r_H} (z+mkr_H^2) \right)
 \nonumber \\
  && \times \exp \left[ {\rm i}mkx -\frac{\gamma}{2r_H^2}(z+mkr_H^2)^2 +\frac{\rm i}{2} (mkr_H)^2 \cot \theta \right],
\end{eqnarray}
where $H_n (w)$ implies the Hermite polynomial. Under a given superposition of LLs, the structure of a vortex lattice is determined by the two parameters $k$ and $\theta$. In the familiar isotropic triangular lattice, we have $k \sqrt{\gamma} = \sqrt{3^{1/2} \pi}/r_H$ and $\theta={\rm tan}^{-1}(\sqrt{3}/\gamma)$. More generally, any basis function $\varphi_n(z,x|-z_0 {\hat z})$ in the gauge ${\bf A}=Hz{\hat x}$ satisfies 
\begin{eqnarray} \label{eq:fomLL1}
  && \varphi_n (z-z_0, x|{\bf 0}) =e^{{\rm i}z_0x/r_H^2} \varphi_n (z, x|-z_0 \hat{z})
 \\ \label{eq:fomLL2}
  && \varphi_n(z,x|-z_0 \hat{z}) =\frac{1}{\sqrt[]{\mathstrut n!}} (\Pi_{+})^n \varphi_0(z,x|-z_0 \hat{z}).
\end{eqnarray}
For later convenience, the following formula \cite{HiasaIkeda} for $\varphi_n(z, x|{\bf 0})$ will be given here : 
\begin{eqnarray}\label{eq:hiasaikeform}
\exp({\rm i}\rho{\bf v}\cdot{\bf \Pi}) \, \varphi_n(z,x|{\bf 0}) &=& \frac{\exp(-|\mu \rho|^2/2)}{\sqrt{n!}} \biggl(\mu^* \rho- \frac{\partial}{\partial (\mu \rho)} \biggr)^n  \nonumber \\
&\times& \exp(\mu^2 \rho^2/2) \, 
\varphi_0(z+ \sqrt{2 \gamma^{-1}} r_H \mu \rho, x|{\bf 0}). 
\end{eqnarray}

Using $\varphi_n(z,x|{\bf 0})$, the "imaginary" term $\propto {\rm sin}(\rho \sigma v_x Q_0)$ in $F^{(s)}_2$ with a nonzero $\delta N$ would induce a coupling between even and odd LLs. To try to exclude such an even-odd coupling and to make LLs better basis functions for diagonalization, the factor $\exp(-{\rm i}\rho \sigma {\bf v}\cdot{\bf \Pi})$ in eq.(\ref{eq:fomROperator}) will be written as $\exp({\rm i}\rho \sigma v_x Q) \exp(-{\rm i}\rho \sigma {\bf v}\cdot{\bf \Pi}_s(Q))$, where ${\bf \Pi}_s(Q)= {\bf \Pi} + Q{\hat x} = -{\rm i}\nabla + r_H^{-2} (z + Q r_H^2) {\hat x}$. Correspondingly, the pair-field is expressed as 
\begin{equation}
\Delta_s = \sum_n Y_{s,n} \, \varphi_n(z+Qr_H^2, x|{\bf 0}).
\end{equation}
Then, using eq.(\ref{eq:hiasaikeform}), the matrix element appearing in $F^{(s)}_2$ is written as 
\begin{equation}
\int d^2r [\varphi_{n_1}(z+Qr_H^2, x|{\bf 0})]^* \exp({\rm i}\rho {\bf v}\cdot{\bf \Pi}_s(Q)) \varphi_{n_2}(z+Qr_H^2, x|{\bf 0}) = \exp(-\rho^2|\mu|^2/2) {\cal L}_{n_1, n_2}(\rho \mu),
\end{equation}
where 
\begin{equation} \label{eq:defLn1n2}
  {\cal L}_{n_1,n_2}(w) =\sum_{n_0=0}^{{\rm min}(n_1,n_2)}
  \frac{\sqrt[]{\mathstrut n_1! n_2!}}{(n_1-n_0)!(n_2-n_0)!n_0!} w^{n_1-n_0} (-w^\ast)^{n_2-n_0}.
\end{equation}
In this way, 
$F^{(s)}_2$ is expressed by 
\begin{eqnarray}\label{eq:Fs2final}
\frac{F^{(s)}_2}{N} &=& \sum_{n_1, n_2} Y^*_{s, n_1} \biggl[ \frac{1}{N w_{ss}} \, \delta_{n_1, n_2} - \int_{\rho_c}^\infty d\rho \, \frac{f(\rho)}{2} \sum_{\sigma=\pm 1} \, \biggl\langle \exp\biggl(-{\rm i}\rho \sigma v_x Q- \frac{\rho^2|\mu|^2}{2} \biggr) \nonumber \\
&\times& |{\hat \Delta}_{\bf k}|^2 ({\rm cos}(\rho v_x Q_0) + {\rm i} \, \delta N \, {\rm sin}(\rho \sigma v_x Q_0)) \, {\cal L}_{n_1, n_2}(\rho \sigma \mu) \, \biggr\rangle \biggr] Y_{s, n_2}.
\end{eqnarray}
As usual, the zero field (mean field) transition temperature $T_c$ can be introduced by deleting $1/(N w_{ss})$ through the relation 
\begin{equation}
\frac{1}{N w_{ss}} = {\rm ln}\frac{T}{T_c} + \int_{\rho_c} d\rho \, \frac{2 \pi T_c}{{\rm sinh}(2 \pi T_c \rho)}. 
\end{equation}
To determine $H_{c2}(T)$ and the pair-field solution giving a free energy minimum at each ($H$, $T$) below $H_{c2}(T)$, we only have to diagonalize the above expression of $F^{(s)}_2$ and to determine $Y_{s,n}$ giving the lowest eigenvalue under a fixed $Q$. 

In the conventional GL region in low fields where both $Q_0$ and ${\bf \Pi}$ are small in magnitude, the imaginary term disappears if identifying $Q$ with $\delta N Q_0$, and $F^{(s)}_2$ is diagonalized via the LL basis functions $\varphi_n(z + Q r_H^2, x|0)$ in the different gauge ${\bf A}=H(z + Q r_H^2){\hat x}$. According to eq.(\ref{eq:fomLL1}), this function is nothing but $\exp(-{\rm i}Qx) \varphi_n(z,x|Q r_H^2 {\hat z})$. Due to the phase factor $e^{-{\rm i}Qx}$, this state was often called a {\it helical vortex state} \cite{KAS}. However, $\varphi_0(z+Qr_H^2, x|{\bf 0})$ itself is an Abrikosov triangular lattice in the lowest LL, and the nonzero $Q$ is {\it not} practically observable in gauge-invariant quantities such as $|\Delta_s({\bf r})|^2$. As shown in our previous report \cite{ymatsu}, the vortex lattices show a single or consecutive first order structural transitions (FOSTs) with increasing field, depending upon the $\delta N$ value, even when assuming $Q=\delta N Q_0$, because the paramagnetic depairing enhanced by increasing the field makes the higher LLs with $n \geq 1$ active. 
However, it is unclear whether or not the phase diagrams obtained under the assumption $Q \simeq \delta N Q_0$ are justified in higher fields. Taking account of this point, we have also examined the phase diagram by directly optimizing the $Q$-value at each ($H$, $T$). As is seen later in the full gap case, however, a direct optimization of the $Q$-value does not significantly change the resulting phase diagram. For this reason, in obtaining phase diagrams for other pairing states, we shall focus later on those following from the relation $Q = \delta N Q_0$. 

So far, we have implicitly assumed that no modulation parallel to ${\bf H} \parallel {\hat y}$ occurs in Rashba superconductors, because the anisotropic Zeeman term in eq.(\ref{eq:QPenergy}) is not accompanied by $k_y$ and hence, does not lead to a paramagnetic depairing effect on the spatial variations parallel to ${\bf H}$ of $\Delta_s$ in contrast to the case of the FFLO state in the lowest LL \cite{AdachiIkeda}. In fact, we have verified this fact concretely in each case of pure singlet pairing. Therefore, throughout this paper, the pair-field can be assumed to be independent of $y$. 

Next, we explain how to evaluate the quartic term of the GL free energy functional. Using the set of $Y_{s,n}$ determined from $F^{(s)}_2$, an equilibrium vortex lattice structure is obtained by minimizing the quartic term. The quartic term $F^{(s)}_4$ is, as well as the last term in eq.(\ref{eq:singletfree}), the sum of contributions from each FS and, in general, takes the form 
\begin{equation} \label{eq:defRF4}
  F^{(s)}_4 =\int d^3{\bf r} \sum_a K_4^{(a)}(\{ {\bf \Pi}_i \}) \Delta_s^\ast({\bf r}_1) \Delta_s({\bf r}_2)
  \Delta_s^\ast({\bf r}_3) \Delta_s({\bf r}_4) \biggr|_{{\bf r}_i \rightarrow{\bf r}}, 
\end{equation}
where 
\begin{eqnarray} \label{eq:defRK4}
  K_4^{(a)}(\{ {\bf \Pi}_i \}) &=& T \sum_{\varepsilon} \int \frac{d^3{\bf k}}{(2\pi)^3}
  {\cal G}_a({\bf k},{\rm i}\varepsilon) {\cal G}_{a}(-{\bf k}+({\bf \Pi}_1^{(a)})^\ast,-{\rm i}\varepsilon)
 \nonumber \\
  && \times {\cal G}_{a}(-{\bf k}+{\bf \Pi}_2^{(a)},-{\rm i}\varepsilon)
  {\cal G}_a({\bf k}+({\bf \Pi}_3^{(a)})^\ast-{\bf \Pi}_2^{(a)},{\rm i}\varepsilon),
\end{eqnarray} 
where ${\bf \Pi}_j^{(a)}=-{\rm i}\nabla_j + 2e {\bf A}({\bf r}_j) + (-1)^a Q_0 {\hat x}$. 
However, when numerous LLs are used in describing $\Delta_s$, it is numerically formidable to exactly examine this expression. For Rashba superconductors, however, replacing the original quartic term with the conventional GL representation 
\begin{equation} \label{eq:defRAbrikosov}
  F^{(s)}_4 \simeq {\tilde F}^{(s)}_4 = c_s \int d^3{\bf r} |\Delta_s({\bf r})|^4
\end{equation}
with a positive coefficient $c_s$ seems to be justified. First, by directly examining $F^{(s)}_4$ for the cases with only the lowest LL and the lowest two LLs, we have found that, in contrast to the centrosymmetric case \cite{AdachiIkeda}, the overall sign of $F^{(s)}_4$ remains positive even in $T \to 0$ limit. Based on these observations, it is believed that $c_s > 0$ so that the first order $H_{c2}$-transition does not occur in noncentrosymmetric superconductors \cite{HI}. Further, for the purpose of determining a stable structure at each ($H$, $T$), even the $H$ and $T$ dependence of $c_s$ is unnecessary. On the other hand, even the nonlocal corrections arising from the orbital depairing have been neglected in $F^{(s)}_4$. Explaining this point will be postponed until our numerical results are presented in this section. 

To represent $|\Delta_s|^2$ in terms of LLs, it is convenient to use the 
formula \cite{RI92}
\begin{equation} \label{eq:fomBilinear}
 \varphi_{n_1}^\ast(z,x|0) \varphi_{n_2}(z,x|{\bf r_0}) = \sum_{\bf G} {\cal L}_{n_2, n_1}\biggl( \frac{(q_x - {\rm i} q_z) r_H}{\sqrt{2}} \biggr) F({\bf G}, {\bf r}_0) e^{{\rm i}{\bf q}\cdot{\tilde {\bf r}}}, 
\end{equation}
where ${\tilde {\bf r}}={\hat z} \sqrt{\gamma} z + {\hat x} x/\sqrt{\gamma}$, 
\begin{equation}
  F({\bf G},{\bf r}_0) =(-1)^{m_1 m_2} {\rm exp} \left[ -\frac{1}{4} {\bf q}^2 r_H^2 + \frac{\rm i}{2}{\bf G}\cdot{\tilde {\bf r}}_0 \right]
\end{equation}
is the Fourier transform of $\varphi_{0}^\ast(z,x|0) \varphi_{0}(z,x|{\bf r}_0)$, ${\bf q}={\bf G}+{\bf k}_0$, ${\bf k}_0=\hat{y}\times{\tilde {\bf r}}_0 /r_H^2$, and ${\tilde {\bf r}}_0 = {\hat z} \sqrt{\gamma} z_0 + {\hat x} x_0/\sqrt{\gamma}$. The reciprocal lattice vector ${\bf G}$ is given by 
\begin{eqnarray} \label{eq:defG}
  {\bf G} &=& m_1{\bf G}_1 +m_2{\bf G}_2, \nonumber \\
  {\bf G}_1 &=& k \sqrt{\gamma} ({\hat x} - {\hat z} \gamma^{-1} {\rm cot}\theta), \nonumber \\
  {\bf G}_2 &=& {\hat z} \, \frac{2 \pi}{k \sqrt{\gamma} r_H^2} 
\end{eqnarray}
under the condition 
\begin{equation}\label{eq:commens}
 N \, |{\bf G}_2| = M \, ({\bf G}_1\cdot{\hat z}), 
\end{equation}
where $m_1$, $m_2$, 
$N$, and $M$ are integers. The condition (\ref{eq:commens}) ensures a periodicity of an obtained vortex lattice in the $x$-direction. Then, using eq.(\ref{eq:fomBilinear}), we have 
\begin{eqnarray} \label{eq:fomAbsRGap}
  |\Delta_s({\bf r})|^2 &=& \sum_{n_1,n_2} Y_{s,n_1}^\ast Y_{s,n_2} [\varphi_{n_1}(z,x|{\bf 0})]^\ast \varphi_{n_2}(z,x|{\bf 0}) \\ \nonumber 
  &=& \sum_{m_1,m_2} (-1)^{m_1m_2} \exp(-|\Gamma|^2/2)
  \sum_{n_1,n_2} Y_{s,n_1}^\ast Y_{s,n_2} {\cal L}_{n_1,n_2}(\Gamma) \, e^{{\bf i}{\bf G}\cdot{\tilde {\bf r}}}, 
\end{eqnarray}
where $\Gamma= - r_H (G_x - {\rm i} G_z)/\sqrt{2}$. Thus, eq.(\ref{eq:defRAbrikosov}) becomes 
\begin{eqnarray}
  \frac{{\tilde F}^{(s)}_4}{V} &=& c_s \sum_{m_1,m_2} e^{-|\Gamma|^2} \Biggl| \sum_{n_1,n_2} Y_{s,n_1}^\ast Y_{s,n_2}
  {\cal L}_{n_1,n_2}(\Gamma) \Biggr|^2,
\end{eqnarray}
and the GL free energy in equilibrium is given by
\begin{equation} \label{eq:REnergy}
  {\cal F}^{(R)} =-\frac{\left( F^{(s)}_2 \right)^2}{2{\tilde F}^{(s)}_4}.
\end{equation}
The equilibrium vortex lattice structure is determined by minimizing ${\tilde F}^{(s)}_4$ or ${\cal F}^{(R)}$ with respect to $k$ and $\theta$. Throughout this paper, we present results of the $h$-$t$ phase diagram obtained in terms of the lowest eight LLs. 

Now, the vortex lattice structures for each pairing state following from the above formulation will be explained. 

\subsubsection{Full gap} 

First, we explain results in the $s$-wave pairing case with no gap nodes where ${\hat \Delta}_{\bf k}=1$. The $\delta N$ dependence of the phase diagram in the $s$-wave case has been previously studied based on the assumption $Q = \delta N Q_0$ \cite{ymatsu}. Here, it will be shown that the previous results are not essentially changed by correctly optimizing the $Q$-value for each field and temperature. 

First, let us start from explaining the important role, mentioned in sec.I, of the anisotropy of the Zeeman energy in the noncentrosymmetric superconductors. As an example, a phase diagram in the full gap case and in $\delta N \to 0$ limit has been shown in Fig.2(b). The two figures in Fig.2 have been obtained by assuming ${\tilde J}=0.2$ and $\mu H_{\rm orb}^{(2D)}(0)/(2 \pi T_c) = 0.4$, where $H_{\rm orb}^{({\rm 2D})}$ is the orbital limiting field at $T=0$ in 2D limit. Hereafter, the $\delta N \to 0$ limit, in which $Q=0$, corresponding to the limit of an infinite band width will be often considered to understand an origin of a strange vortex lattice structure, and the applied field value will be denoted as $h=H/H_{\rm orb}^{({\rm 2D})}(0)$. In Fig.2(b), intermediate lattice structures changing with varying the field appear. They are induced by higher even LLs contributions to $\Delta_s$ with $n=2m$ ($m > 0$), which play enhanced roles in noncentrosymmetric systems through the $v_x \propto {\hat k}_x$ dependence in the Zeeman energy term of eq.(\ref{eq:NCSGreenfunc}). Such intermediate phases do not appear in Fig.2(a) corresponding to the familiar centrosymmetic case which is obtained by replacing ${\rm cos}(\rho v_x Q_0)$ in eq.(\ref{eq:Fs2final}) with ${\rm cos}(\rho v_F Q_0)$ independent of ${\hat {\bf k}}$ and setting $\delta N=Q=0$. Thus, the intermediate states in Fig.2(b) can occur even in Rashba superconductors with smaller Maki parameters where the high field state (A) of Larkin-Ovchinnikov (LO) type does not appear. Further, the states (B) and (C) in Fig.2(b) and the continuous crossover between them are also seen in the singlet-triplet ($s$-$t$) mixed case to be discussed later.

\begin{figure}[t]
\caption{(Color online) Resulting $h$-$t$ phase diagrams for (a) $\delta N=0$, (b) $\delta N=-0.003$, and (c) $\delta N=-0.2$ and (d) $-Q/Q_0$ v.s.$h$ curve taken on $H_{c2}(T)$ of the figure (c) in the full gap ($s$-wave pairing) case, where $h = H/H_{\rm orb}^{({\rm 2D})}(0)$ and $t=T/T_c$. With increasing $|\delta N|$, the $H_{c2}$-value merely shows a slight increase, while the corresponding change of the vortex lattice structure is drastic. The intermediate structures with no reflection symmetry, seen in Fig.2(b), compete with the novel striped structure, (B) of (b), modulating along ${\hat x}$ perpendicular to both of ${\bf H}$ and the $c$-axis and are overcome by the latter at higher $\delta N$-values. The parameter values $\mu H_{\rm orb}^{(2D)}(0)/(2 \pi T_c) = 0.4$, ${\tilde J}=0.1$, and $\rho_c = (20 \pi T_c)^{-1}$ are commonly used. } \label{fig.3}
\end{figure}

In Fig.3, our results in the $s$-wave case obtained by directly optimizing the $Q$-value are shown together with the resulting $Q(h)$ data obtained along the $H_{c2}(T)$-curve in $\delta N = 0.2$ case. All figures in this section including Fig.3 have been obtained by using ${\tilde J}=0.1$. The $H_{c2}(T)$-value is slightly enhanced with increasing $|\delta N|$ and thus, $|Q| \propto |\zeta|$, suggesting that an increase of spin-orbit coupling diminishes the paramagnetic depairing. The resulting phase diagram in $\delta N=0$ case, Fig.3(a), is quite similar to Fig.2(b) (see, however, the next paragraph). By assuming a very small but nonvanishing $\delta N$,  not only the LO type state, (A) in Fig.3(a), but also the intermediate states (B) and (C) there are destabilized by the appearance of the {\it novel} modulated state, (B) and (C) in Fig.3(b), which have reflection symmetry in contrast to (B) and (C) in Fig.3(a) and are stabilized by the {\it hidden} ${\bf Q}=Q {\hat x}$-vector or the resulting anisotropy (see sec.I). Throughout this paper, we often encounter a compressed square lattice in Rashba superconductors, which is created through a field-induced crossover from the stripe-like modulated state, (B) in Fig.3 (b). As $\delta N$ is increased further, the states (A), (B), and (C) in Fig.3(a) with {\it no} reflection symmetry in the plane perpendicular to ${\bf H}$ are completely lost, and the high field state for $\delta N=0.2$ is the square lattice corresponding to (C) of Fig.3(b) which continuously occurs through a crossover from (B) in Fig.3(b). Further, even the triangular vortex lattice at higher temperatures is compressed with increasing $|\delta N|$, possibly reflecting mixings between even and odd LLs induced by a nonzero $\delta N$, and the only structural transition surviving at higher $|\delta N|$ is the FOST between the compressed square and triangular lattices. Note that there are two types of continuous crossovers in structure in intermediated fields. As well as the crossover expressed as (B) and (C) in Fig.2(b), this crossover between (B) and (C) in Fig.3(b) also appears in the more general case with $s$-$t$ mixing of pairing channels. 

Here, the difference in the orientation of the low field triangular lattice between Fig.3(a) and other figures in Fig.2 and 3 will be commented on. First, in Fig.3, we find that even a small but nonvanishing $|\delta N|$ changes the orientation of the triangular lattice, and that, at a fixed (and nonvanishing) $\delta N$, there is no indication of an orientational transition in the triangular lattice region. Below the dashed line in Fig.3(a), however, the orientation is found to be the same as in Fig.3(c). This feature is commonly seen in the cases of other pairing states to be given below. On the other hand, the difference in the orientation between Fig.2(b) and Fig.3(a) can be attributed to the difference in the ${\tilde J}$-value used in calculations, or equivalently, to the strength of the paramagnetic depairing: As the paramagnetic depairing is enhanced, higher LL modes with {\it even} indices in the pair-field become more effective and change the orientation. This statement was justified by separately performing a calculation taking account only of the lowest LL. Besides, the absence of such an orientational change in the case with a nonvanishing $\delta N$ is also consistent with the effective reduction of the paramagnetic depairing due to a nonzero $\delta N$ mentioned in the preceding paragraph. Therefore, as far as the uniaxial anisotropy measured by $\gamma$ is not so large, the orientation of the low field triangular lattice in real systems with nonvanishing $|\delta N|$ is expected to keep that of Fig.3(c). Hereafter, we will not discuss this possibility of an orientational transition any longer. 

Quantitatively, there are some differences between the present Fig.3 following from the $Q$-optimization and Fig.3 in Ref.\cite{ymatsu} where $Q= \delta N Q_0$ was assumed. For instance, in the latter for $\delta N=0.003$, the LO-like state, (A) in the present Fig.3(a) and (b), survives over a broader field range compared with that in the corresponding one in the former. However, except such quantitatively subtle differences, there were no notable differences due to the $Q$-value in the phase diagrams and their $\delta N$ dependences. This evidently shows that the analysis in Ref.\cite{ymatsu} trying to take care of the approximation on the $Q$-value by including the eight LLs is justified. Further, judging from this fact that the vortex structure is not sensitive to the $Q$-value, our neglect of the $k_z$-dependence accompanying $Q_0$ in eq.(20) is also believed to be safely valid. 

Here, based on the figures in Fig.3, our replacement of the original quartic term $F^{(s)}_4$ with the conventional local expression ${\tilde F}^{(s)}_4$ will be discussed. This replacement is safely valid in centrosymmetric superconductors with a large paramagnetic effect \cite{HM,KM}. Even in the present Rashba case, the same thing should hold. For smaller $|\delta N|$, the paramagnetic effect is stronger, and intriguing modulated vortex structures tend to appear. Although there might be possibility that the validity of this local approximation is subtle in intermediate fields, the structural changes in the case with nonzero $\delta N$ are smooth so that most of FOSTs in $\delta N=0$ case are changed into crossovers. In such a crossover regime, a large deviation between the results of $F^{(s)}_4$ and of ${\tilde F}^{(s)}_4$ is not expected. For this reason, we believe that the use of the local approximation for $F^{(s)}_4$ is qualitatively valid and does not lead to a significant error in our results on the vortex lattice structure. 

\begin{figure}[b]
\caption{(Color online) The resulting $h$-$t$ phase diagrams for (a) $\delta N=0$ and (b) $\delta N=-0.1$ in the case with ${\hat \Delta}_{\bf k}=\sqrt{2} {\rm cos}(2{\tilde k}_z)$. Other parameter values are the same as in Fig.3.} \label{fig.4}
\end{figure}

\subsubsection{Horizontal line nodes} 

Here, phase diagrams in the case of a superconducting gap with horizontal line nodes will be briefly explained. Recently, such a nodal gap has been proposed as a model of CeRhSi$_3$ and CeIrSi$_3$ \cite{Tada}, and, following Ref.\cite{Tada}, we choose the gap function ${\hat \Delta}_{\bf k}=\sqrt{2} {\rm cos}(2 {\tilde k}_z)$. Figure 4 includes the resulting phase diagrams in this case. 
Clearly, the obtained phase diagrams are essentially the same as those in the full gap case. This is due to the fact that, in the angular average over each FS, the ${\tilde k}_z$ dependence of the gap function does not directly couple to the ${\hat k}_x$ dependence in the Zeeman energy. Thus, the same thing should hold for any gap function dependent only on ${\tilde k}_z$. Therefore, it appears that, when the $s$-$t$ mixing is negligible, the presence of a horizontal line node in the superconducting gap cannot be judged from the resulting vortex lattice structure. 

\subsubsection{Vertical line nodes} 

In contrast to the preceding case, the momentum dependence in the gap function directly couples to that of the Zeeman term when the gap nodes consist of vertical lines, leading to a drastic change of vortex lattice structure. This will be explained here in the $d_{x^2-y^2}$-pairing case where ${\hat \Delta}_{\bf k}=\sqrt{2} {\rm cos}(2 \phi_{\bf k})$.

\begin{figure}[b]
\caption{(Color online) The resulting $h$-$t$ phase diagrams in the $d_{x^2-y^2}$-pairing case in magnetic fields applied along (a) a node and (b) an antinode of the energy gap. For both figures, $\delta N=0$ was assumed, and other parameter values are the same as those in Fig.3. Note the modulation parallel to the $c$-axis and the absence of intermediate phases in (a).} \label{fig.5}
\end{figure}

In this $d_{x^2-y^2}$-pairing case, situation changes depending on whether the applied field ${\bf H}$ is along the nodal direction or antinodal one. Differences between these two cases already appear in the $H_{c2}$-curves shown in Fig.5. In ${\bf H}$ parallel to an antinode, the $H_{c2}(T)$ value is slightly enhanced compared with that in the full gap case obtained in terms of the same set of parameter values. In contrast, $H_{c2}(T)$ in ${\bf H}$ parallel to a gap node is remarkably depressed compared with the full gap curve. This strong anisotropy in $H_{c2}$ is a consequence of the coupling in the momentum dependence between ${\hat \Delta}_{\bf k}$ and the anisotropic Zeeman term. 

A more remarkable difference is seen in the resulting vortex lattice structures in the two field configurations. In ${\bf H}$ parallel to an antinode, the vortex lattice structure is qualitatively the same as in the preceding two cases, while, in ${\bf H}$ parallel to a gap node, there are no intermediate states in $\delta N=0$ limit, and the high field state is of the LO-type with an unidirectional modulation parallel to ${\hat z}$, i.e., the $c$-axis. In this case, the direction of the modulation in the LO-like state is pinned by the vertical line nodes. The resulting vortex lattices always have reflection symmetry in the plane perpendicular to ${\bf H}$ in contrast to the intermediate phases in Fig.3(a) and Fig.4(a). This fact may have crucial consequences in the cases with a significant mixing of a $d$-wave component with vertical line nodes and the corresponding $f$-wave one (see sec.V). 

\begin{figure}[t]
\caption{(Color online) The resulting $h$-$t$ phase diagram in the $d_{x^2-y^2}$-pairing system with $\delta N=-0.003$ in magnetic fields applied along a node. The same parameters are used as in Fig.5} \label{fig.6}
\end{figure}

To understand whether this {\it vertical} LO state survives for realistic $\delta N$ values, we have also examined the corresponding phase diagrams for nonzero $\delta N$ values and have found that, for $|\delta N| > 0.1$, the vertical LO state is absent even close to $H_{c2}(0)$. As Fig.6 shows, however, the novel striped state with modulation perpendicular to the $c$-axis, corresponding to (B) in Fig.3(b), occupies a much narrower region compared with that in Fig.3(b) in the full gap case, because the LO-like state competing with this novel intermediate state is supported in this case by the vertical line nodes. Thus, we expect that the region in which the novel modulated state replaces the vertical LO-like state is narrower even for more realistic $\delta N$ values, and thus that the LO-like state may be observable as a high field state in this case in contrast to that in the preceding two cases. For these reasons, we believe that the presence of vertical line gap-nodes can be anticipated by investigating vortex lattice structures.

\section{IV Singlet-Triplet Mixed case}

\subsubsection{Quasi 2D case}

In the preceding section, the singlet ($s$)-triplet ($t$) mixing, which is usually present in noncentrosymetric superconductors with nonzero $\delta N$, has been neglected by assuming one of the two channels to be dominantly attractive. This approximation is valid for a vanishingly small $|\delta w|$, i.e., as far as one of $w_{ss}$ and $w_{tt}$ is small enough (see eq.(\ref{eq:deltaw})). However, when the ratio $w_{tt}/w_{ss}$ is of order unity, this $s$-$t$ mixing drastically changes the $H$-$T$ phase diagram even if $|\delta N|$ is vanishingly small. That is, the $s$-$t$ mixing is measured by a finite $w_{tt}/w_{ss}$, i.e., eq.(\ref{eq:deltaw}), in nonzero fields rather than a nonzero $\delta N$ \cite{FAKS} in zero field case, as a result of the fact that the pair-field in nonzero fields is intrinsically spatially varying. 

In this section, roles of the $s$-$t$-mixing will be first examined in details for quasi two-dimensional (Q2D) systems with a large $\gamma$. For simplicity, we focus on the case of a mixing of $s$-wave and $p$-wave pairings. In this Q2D case, the momentum dependence of $|{\bf g}_{\bf k}|$ appearing in some places may be neglected to simplify our evaluation of the free energy. To clarify the details of this treatment, let us first start from introducing the pair-field on each of the split FSs in the $s$-$t$ mixed case. As the expressions of pair-field operators $\Psi^{(j)}_{\bf p}$ ($j=s$ and $t$) suggest, the (spatially varying) energy gap on the FS $a$ is generally given by 
\begin{equation}\label{eq:deltaa}
  \Delta_a =\frac{\Delta_s -(-1)^a |{\bf g}_{\bf k}| \Delta_t}{\sqrt[]{\mathstrut 2}} \quad (a=1,2)
\end{equation}
which is accompanied by the momentum dependence ${\bf g}_{\bf k}$ of the spin-orbit coupling even after having been separated from ${\hat \Delta}_{\bf k}$-dependence, where $\Delta_s$ ($\Delta_t$) is the singlet (triplet) gap corresponding to $\Psi^{(s)}$ ($\Psi^{(t)}$). In Q2D case, the factor $|{\bf g}_{\bf k}|$ in $\Delta_a$ is replaced by unity, and 
\begin{equation}\label{eq:Q2Ddelta}
\Delta_a \simeq \Delta_a^{(0)} \equiv \frac{\Delta_s 
-(-1)^a \Delta_t}{\sqrt{2}}
\end{equation}
will be used in this section. 

Derivation of the quadratic GL term $F^{({\rm Q2D})}_2$ is almost the same as in the preceding section once $\Delta_j$ ($j=s$ and $t$) are expressed via $\Delta_a^{(0)}$ through eq.(\ref{eq:Q2Ddelta}), and we obtain 
\begin{eqnarray} \label{eq:defRF2}
  F^{({\rm Q2D})}_2 &=&\int d^3{\bf r} \biggl[ \sum_a \biggl( \biggl[ \frac{(w^{-1})_{ss} +(w^{-1})_{tt}}{2} -(-1)^a (w^{-1})_{st}
  \biggr] |\Delta_a^{(0)}|^2 - 2 (\Delta_a^{(0)})^\ast K_2^{(a)}({\bf \Pi}) \Delta_a^{(0)} \biggr)
 \nonumber \\
  && + \biggl[ \frac{(w^{-1})_{ss}-(w^{-1})_{tt}}{2} \biggr] \bigl[ \Delta_1^\ast \Delta_2 +({\rm c.c.}) \bigr] \biggr].
\end{eqnarray}
According to the expression (\ref{eq:defRKs}) of $K_2^{(a)}$, it is natural in this case to choose the gauge in the manner depending on each FS and to 
represent the pair-field in the form
\begin{equation} \label{eq:defRGap2}
  \Delta_a^{(0)} =\sum_{n \geq 0} Y_{a,n} \varphi_n ({\bf r}_a |{\bf 0}),
\end{equation}
with ${\bf r}_a={\bf r}+(-1)^a Q_0 r_H^2 \hat{z}$. Note that the gauge-invariant gradient corresponding to $\varphi_n({\bf r}_a|{\bf 0})$ is 
\begin{equation}
{\bf \Pi}_a = -{\rm i}\nabla + r_H^{-2}(z + (-1)^a Q_0 r_H^2){\hat x}. 
\end{equation}
Using the formula (\ref{eq:hiasaikeform}), the term including the $\rho$-integral in eq.(\ref{eq:defRF2}) becomes 
\begin{equation} \label{eq:defRK2}
  \int d^3{\bf r} \, (\Delta_a^{(0)})^\ast K_2^{(a)}({\bf \Pi}) \Delta_a^{(0)} =N_a V\sum_{n_1,n_2} Y_{a,n_1}^\ast Y_{a,n_2}
  \int_{\rho_c}^{\infty} d\rho f(\rho) \Bigl\langle e^{-\frac{1}{2}|\mu|^2 \rho^2}
  {\rm Re} \, {\cal L}_{n_1,n_2}(\mu\rho) \Bigr\rangle. 
\end{equation}

To calculate the off-diagonal (last) term in eq.(\ref{eq:defRF2}), we will directly use the formula (\ref{eq:fomBilinear}), and consequently, 
\begin{eqnarray}
  \int d^3{\bf r} (\Delta_1^{(0)})^\ast \Delta_2^{(0)} &=& \sum_{n_1,n_2}Y_{1,n_1}^\ast Y_{2,n_2} \int dzdx
  \varphi_{n_1}^\ast({\bf r}_1|{\bf 0}) \varphi_{n_2}({\bf r}_1|2Q_0 r_H^2\hat{z}) e^{-{\rm i}2 Q_0 x} \\ \nonumber 
   &=& V \sum_{n_1,n_2}Y_{1,n_1}^\ast Y_{2,n_2}
  e^{-\gamma Q_0^2 r_H^2} {\cal L}_{n_2,n_1}(\ \sqrt[]{\mathstrut 2\gamma}Q_0 r_H), 
\end{eqnarray}
where the property (\ref{eq:fomLL1}) was used. 

Therefore, eq.(\ref{eq:defRF2}) takes the form
\begin{eqnarray} \label{eq:RF2}
  \frac{F^{({\rm Q2D})}_2}{V} &=& \sum_{a,n_1,n_2} Y_{a,n_1}^\ast Y_{a,n_2} \biggl[ \left( \frac{(w^{-1})_{ss}+(w^{-1})_{tt}}{2} 
  -(-1)^a (w^{-1})_{st} \right) \delta_{n_1,n_2} 
 \nonumber \\
  && \quad - 2 N_a \int_{\rho_c}^\infty d\rho f(\rho) \Bigl\langle e^{-\frac{1}{2}|\mu|^2 \rho^2}
  {\rm Re}{\cal L}_{n_1,n_2}(\mu\rho) \Bigr\rangle  \biggr]
 \nonumber \\
  &&- N (\delta w)^{-1} \sum_{n_1,n_2} \left( Y^\ast _{1,n_1}Y_{2,n_2} e^{-\gamma Q^2r_H^2}
  {\cal L}_{n_2,n_1}(\ \sqrt{\mathstrut 2\gamma}Qr_H) +({\rm c.c.}) \right). 
\end{eqnarray}
Here, it is important to note that the paramagnetic effect appears through the FFLO wavenumber $Q_0$ only in the last term, i.e., the cross term between different FSs. Thus, in $|\delta w| \to \infty$ limit, i.e., when $w_{ss} \simeq w_{tt}$ (see eq.(\ref{eq:deltaw})),  
the paramagnetic effect is lost irrespective of the $w_{st}$-value, and the orbitally-limited situation is realized. The parameter $|\delta w|$ measures the magnitude of the $s$-$t$ mixing in {\it nonzero} fields. In the present Q2D model, the energy gap on the FS with a smaller density of states vanishes (see, for instance, Fig.8 shown below) in $|\delta w| \to \infty$ 
limit. 
The single-pairing case, that is, the pure singlet or pure triplet case corresponds to the case with vanishing $\delta w$, while the single and triplet channels  equally contribute and are competitive with each other when $|\delta w| \to \infty$. 

The zero field transition temperature $T_c$ is given by $w_{ij}$ through the expression 
\begin{equation}
  \frac{(w^{-1})_{ss}+(w^{-1})_{tt}}{2N} =\int_{\rho_c}^\infty d\rho f(\rho)|_{T=T_c} 
  +\sqrt[]{\mathstrut (\delta w)^{-2} +\left( \delta N \int_{\rho_c}^\infty d\rho f(\rho)|_{T=T_c}
  + \frac{(w^{-1})_{st}}{N} \right)^2}. 
\end{equation}
It is not difficult to verify that the above expression is equivalent to eqs.(28) and (31) in the impurity-free case in Ref.\cite{Frigeri06}.

Regarding the quartic term, the method in the single pairing case will be directly used. The quartic term $F^{({\rm Q2D})}_4$ is given by eq.(\ref{eq:defRF4}) with $\Delta_s$ replaced by $\Delta_a^{(0)}$, and, for a similar reason to that in the single pairing case, it will be replaced by its local expression 
\begin{equation} \label{eq:defRAbrikosov}
  F^{({\rm Q2D})}_4 \simeq c_2 \int d^3{\bf r} \sum_a N_a 
|\Delta_a^{(0)}({\bf r})|^4,
\end{equation}
where $c_2$ is positive.  Then, the quartic term to be used for determining the lattice structure is given by 
\begin{eqnarray}
  \frac{F^{({\rm Q2D})}_4}{V} &=& c_2 \sum_a N_a \sum_{m_1,m_2} e^{-|\Gamma|^2} \Biggl| \sum_{n_1,n_2} Y_{a,n_1}^\ast Y_{a,n_2}
  {\cal L}_{n_1,n_2}(\Gamma) \Biggr|^2,
\end{eqnarray}
and the GL free energy in equilibrium is given by
\begin{equation} \label{eq:REnergy}
  {\cal F}^{(R)} =-\frac{\left( F^{(R)}_2 \right)^2}{2F^{(R)}_4}.
\end{equation}

\begin{figure}[t]
\caption{(Color online) The resulting $h$-$t$ phase diagram in the $s$-wave and $p$-wave mixed case based on the Q2D approximation (\ref{eq:Q2Ddelta}). The used parameters are ${\tilde J}=0.2$, $\delta N=0$ and $\delta w=1.25$. For each point in the phase diagram, the four images in real space of $|\Delta_1|$, $|\Delta_2|$, $|\Delta_s|$, and $|\Delta_t|$ are shown from top to bottom.} \label{fig.7}
\end{figure}

We show, in Figs.7 and 8, examples of the resulting phase diagrams and vortex lattice structures in the case with a mixing of $s$-wave and $p$-wave components {\it and} with a Q2D (cylindrical) FS. In those figures, the vortex lattice structures at several selected points in each $h$-$t$ phase diagram are represented as real space patterns of $|\Delta_1^{(0)}|$ on FS1, $|\Delta_2^{(0)}|$ on FS2, $|\Delta_s|$, and $|\Delta_t|$. In general, when $|\delta N|$ is essentially 
zero so that both of the two FSs contribute equally to superconductivity, as shown in Fig.7, the vortex positions in $\Delta_1^{(0)}$ do not coincide with those in $\Delta_2^{(0)}$, reflecting the fact that the spatial pattern of $|\Delta_s|$ tends to become quite opposite to that of $|\Delta_t|$. In this case, when $\Delta_s$ consists of even LLs, $\Delta_t$ is expressed only by the odd LLs, and vice versa. On the other hand, with increasing $|\delta N|$, one of the two FSs dominantly contributes to superconductivity, and, according to eq.(\ref{eq:Q2Ddelta}), the (nearly) zero points of $|\Delta_s|$ coincide with those of $|\Delta_t|$ (see Fig.8). When one of the two FSs primarily determines superconductivity, the paramagnetic depairing is significantly reduced, because the paramagnetic effect on {\it a single} FS is trivially gauged away \cite{HI} (see also Fig.1(b)). As in the case with a single pairing component, an increase of $|\delta N|$ results in a significant mixing between the even and odd LLs, and, as in Fig.8, $|\Delta_s|$ tends to show similar spatial patterns to $|\Delta_t|$. 

\begin{figure}[b]
\caption{(Color online) The resulting $h$-$t$ phase diagram in the $s$-wave and $p$-wave mixed case based on the Q2D approximation (\ref{eq:Q2Ddelta}). The used parameters are ${\tilde J}=0.2$, $\delta N=-0.1$ and $\delta w=5$. } \label{fig.8}
\end{figure}

Figure 7 corresponds to a phase diagram of the case with a slight inclusion of a $p$-wave pairing component in the pure $s$-wave case. Here, $\delta N$ is set to be zero, and hence, this figure is comparable with Fig.3 (a) and (b). Similarly to the difference between Fig.3(a) and Fig.3(b) induced by a slight increase of $|\delta N|$, the intermediate state (C) in Fig.7 with no reflection symmetry are limited to a narrow region and dominated by the stripe-like modulated lattice (A) and (B). However, the structure at (B) in Fig.7 is not a square lattice appearing as a crossover from the striped structure (A) but rather a triangular lattice which can be obtained by rotating another triangular lattice appearing in lower fields just below FOST4. The triangular lattice in intermediate fields is also present even for larger $|\delta w|$, i.e., even when the $s$ and $p$-wave components are more significantly mixed, while it is lost as $|\delta N|$ is larger, as can be seen in Fig.8 where a more $s$-$t$ mixing and a larger $|\delta N|$ than in Fig.7 were assumed. A typical $\delta N$ dependence of $H_{c2}(T)$-curve including that of Fig.8 has been given in Ref.\cite{HI}. In the $\delta N=0$ limit, the $h$-value corresponding to $H_{c2}(0)$ is close to $3.0$. Such a much larger $H_{c2}(0)$ than that of Fig.3(a) is due to the $s$-$t$ mixing, which clearly plays more dominant roles than a nonzero $\delta N$ for enhancing $H_{c2}$. Nevertheless, Fig.8 has similar features to those of Fig.3(c). For instance, in both Fig.3(c) and Fig.8, the triangular lattice near $T_c$ shows an anisotropic structure compressed along the $c$-axis, while the vortex structure in higher fields is an anisotropic square lattice created from the novel striped lattice (B) in Fig.3(b). Therefore, in the present case with a cylindrical FS with a negligibly small corrugation, an increase of a $s$-$t$ mixing plays qualitatively similar roles to an increase of the magnitude of the spin-orbit coupling $|\delta N|$ in the $h$-$t$ phase diagram, and, in a realistic situation where both $\delta w$ and $|\delta N|$ are nonvanishing, vortex lattice structures with no reflection symmetry such as (B) in Fig.3(a) are expected not to occur. 

In (C) of Fig.8, the image of $|\Delta_s|$ is much brighter than that of $|\Delta_t|$, while both of them in (A) are almost the same as each other. The former feature in lower fields is a reflection of the fact that, in zero field, the singlet component is the dominant pairing state, and a small $|\delta N|$ induces the triplet componet, while, with increasing field, the role of inducing a $s$-$t$ mixing is played not by $|\delta N|$ but rather by the nonvanishing $\delta w$. In particular, at high enough fields and in low temperature limit, the vortex structure is an anisotropic square lattice oriented along the $c$ axis for any $\delta N$, implying that the phase diagram there is sensitive not to $\delta N$ but to $|\delta w|$. In the next subsection, however, these conclusions in Q2D case are found to be changed for more three dimension(3D)-like FSs. 
For instance, 
the $h$-$t$ phase diagram for a more 3D-like FS seems to have a much stronger $\delta N$ dependence than that seen above. 

In obtaining Fig.8, we have assumed $\delta N < 0$ and $w_{st} > 0$. Since $w_{st}$ generally depends on the higher energy cutoff, the results following from diagonalization of $F^{(Q2D)}_2$ are quantitatively affected by the details of $w_{st}$. In fact, if $w_{st}$ is zero or negative, $|\Delta_1|$ rather than $|\Delta_2|$ should be larger. The opposite result to this, seen in Fig.8, is a consequence of a positive $w_{st}$. However, we have reexamined Fig.8 by changing the sign of $w_{st}$ and have found that the field induced changes of the vortex lattice structure remain qualitatively the same as those in Fig.8. Based on this fact, we shall assume hereafter the vortex lattice structure to be qualitatively insensitive to the sign of $w_{st}$.

\subsubsection{More 3D-like case}

In this subsection, the Q2D approximation in the last section is relaxed, and effects of the corrugation parallel to the $c$-axis of the Q2D Fermi surface will be incorporated. Then, the $k_z$-dependence in $\Delta_a$ neglected in the last section needs to be included. In contrast to the weak $k_z$ dependence accompanying the parameter $Q_0$, neglected for simplicity in eq.(20), this $k_z$ dependence may lead to a change of the degree of mixing of even and odd LLs which affects the vortex lattice structure. To simplify our treatment, the factor $|{\bf g}_{\bf k}|$ in $\Delta_a$ will be approximated by 
\begin{eqnarray}
  |{\bf g}_{\bf k}| &=& \sqrt[]{\mathstrut 1-\tilde{J}(1-{\rm cos}k_z)}
 \\
  &\simeq & \sqrt[]{\mathstrut 1-\tilde{J}} +\frac{\tilde{J} {\rm cos}k_z}{2\ \sqrt[]{\mathstrut 1-\tilde{J}}}.
\end{eqnarray}
Then, using 
\begin{equation}
  {\tilde \Delta}_a^{(0)} =\frac{\Delta_s -(-1)^a\ \sqrt[]{\mathstrut 1-\tilde{J}} \Delta_t}{\sqrt[]{\mathstrut 2}},
\end{equation}
$\Delta_a$ is expressed as 
\begin{eqnarray} \label{eq:CPSGap}
  \Delta_a &=& {\tilde \Delta}_a^{(0)} -(-1)^a \frac{\tilde{J} {\rm cos}k_z}{2\ \sqrt[]{\mathstrut 2(1-\tilde{J})}} \Delta_t
 \nonumber \\
  &=& (1+\varsigma){\tilde \Delta}^{(0)}_a-\varsigma {\tilde \Delta}^{(0)}_b,
\end{eqnarray}
where $a$ and $b=1$ or $2$, $a \neq b$, and 
\begin{equation}
  \varsigma = \frac{\tilde{J}{\rm cos}k_z}{4(1-\tilde{J})}.
\end{equation}

First, let us start from rewriting terms dependent on $w_{ij}$ in the GL quadratic terms (see eq.(\ref{eq:defRF2})) into the form 
\begin{eqnarray} \label{eq:CPSF2first}
  2 \, \sum_{i,j=s,t} \Bigl\langle (w^{-1})_{ij} \Delta_i^\ast \Delta_j \Bigr\rangle &=&  [(w^{-1})_{ss}+(\tilde{w}^{-1})_{tt}+2(\tilde{w}^{-1})_{st}] |{\tilde \Delta}^{(0)}_1|^2
  +[(w^{-1})_{ss}+(\tilde{w}^{-1})_{tt}-2(\tilde{w}^{-1})_{st}] |{\tilde \Delta}^{(0)}_2|^2 
 \nonumber \\
  && + 2 [(w^{-1})_{ss}-(\tilde{w}^{-1})_{tt}]{\rm Re}({\tilde \Delta}^{(0)\ast}_1 {\tilde \Delta}^{(0)}_2) 
\end{eqnarray}
expressed by ${\tilde \Delta}^{(0)}_a$, where $(\tilde{w}^{-1})_{tt} = (w^{-1})_{tt}/(1-\tilde{J})$, and $(\tilde{w}^{-1})_{st} = (w^{-1})_{st}/\sqrt{1-\tilde{J}}$. 
The remaining term including the kernel $K_2^{(a)}$ is given by
\begin{eqnarray} \label{eq:CPSF2second}
  -2 \, \sum_a \Bigl\langle \Delta_a^\ast \, K_2^{(a)}({\bf \Pi}) \, \Delta_a \Bigr\rangle &=& -2\int_{\rho_c}^\infty d\rho f(\rho)
  \sum_{s_\varepsilon,a} N_a \Bigl\langle \Delta_a^\ast e^{-s_\varepsilon{\rm i}\rho{\bf v}\cdot{\bf \Pi}_a} \Delta_a
  \Bigr\rangle
 \nonumber \\
  &=& -2 \int_{\rho_c}^\infty d\rho f(\rho)\sum_{s_\varepsilon,a} N_a
 \nonumber \\
  && \times \Bigl\langle (1+\varsigma)^2 ({\tilde \Delta}^{(0)}_a)^\ast e^{-s_\varepsilon{\rm i}\rho{\bf v}\cdot{\bf \Pi}_a}
  {\tilde \Delta}^{(0)}_a +\varsigma^2 e^{s_{\varepsilon}{\rm i}\rho 2Q_0 v_x} ({\tilde \Delta}^{(0)}_b)^\ast 
  e^{-s_\varepsilon{\rm i}\rho{\bf v}\cdot{\bf \Pi}_b} 
{\tilde \Delta}^{(0)}_b
 \nonumber \\
  && \quad -\varsigma(1+\varsigma) \left( ({\tilde \Delta}_a^{(0)})^\ast e^{-s_{\varepsilon}{\rm i}\rho{\bf v}\cdot{\bf \Pi}_a}
  {\tilde \Delta}^{(0)}_b +({\tilde \Delta}^{(0)}_b)^\ast e^{-s_{\varepsilon}{\rm i}\rho{\bf v}\cdot{\bf \Pi}_a} 
  {\tilde \Delta}^{(0)}_a \right) \Bigr\rangle.  
\end{eqnarray}
Note that this term is diagonal with respect to $\Delta_a$, while it is not diagonalized any longer in the ${\tilde \Delta}^{(0)}_a$-representation. Then, by representing ${\tilde \Delta}_a^{(0)}$, as in eq.(\ref{eq:defRGap2}), in the form
\begin{equation} \label{eq:CPSGap0}
  {\tilde \Delta}^{(0)}_a =\sum_{n\geq 0} Y_{a,n} \varphi_n ({\bf r}_a|0),
\end{equation}
the cross term between ${\tilde \Delta}^{(0)}_b$ and ${\tilde \Delta}^{(0)}_a$ in eq.(\ref{eq:CPSF2second}) becomes 
\begin{equation} \label{eq:CPSF2CrossTerm}
  \int d^3 {\bf r} ({\tilde \Delta}^{(0)}_b)^\ast e^{-s_\varepsilon{\rm i}\rho{\bf v}\cdot{\bf \Pi}_a} {\tilde \Delta}^{(0)}_a 
  =\sum_{n_1,n_2} Y_{b,n_1}^\ast Y_{a,n_2} J_{n_1,n_2}\left( (-1)^b\sqrt[]{\mathstrut 2\gamma}Q_0 r_H, s_\varepsilon\mu\rho
  \right), 
\end{equation}
where 
\begin{eqnarray} \label{eq:defJn1n2}
  && J_{n_1,n_2}\left( (-1)^b\sqrt[]{\mathstrut 2\gamma}Q_0 r_H ,s_\epsilon\mu\rho \right) =\int d^3 {\bf r}
  \varphi_{n_1}^\ast({\bf r}_b|0) e^{-s_\varepsilon{\rm i}\rho{\bf v}\cdot{\bf \Pi}_a} \varphi_{n_2}({\bf r}_a|0)
 \nonumber \\
  && =\frac{1}{\sqrt{\mathstrut n_2!}} e^{-\frac{1}{2}|\mu|^2\rho^2} \left( -s_\varepsilon \mu^\ast\rho
  +s_\varepsilon \frac{\partial}{\partial(\mu\rho)}\right)^{n_2} e^{\frac{1}{2}\mu^2\rho^2} \int d^2{\bf r}
  \varphi_{n_1}^\ast({\bf r}_b|0)\varphi_0({\bf r}_a -s_\varepsilon\ \sqrt{\mathstrut 2}r_H\mu\rho\hat{z}|0)
 \nonumber \\
  && =\frac{1}{\sqrt{\mathstrut n_1!n_2!}} e^{\gamma Q_0^2 r_H^2-\frac{1}{2}|\mu|^2\rho^2}
  \left( -s_\varepsilon \mu^\ast\rho +s_\varepsilon \frac{\partial}{\partial(\mu\rho)} \right)^{n_2}
  \left( (-1)^b \sqrt{\mathstrut 2\gamma}Q_0 r_H +s_\varepsilon\mu\rho \right)^{n_1}
  e^{-2\gamma Q_0^2 r_H^2-(-1)^b s_\varepsilon\ \sqrt{\mathstrut 2\gamma} Q_0 r_H \mu \rho}
 \nonumber \\
  && =e^{-\frac{1}{2}|\mu|^2\rho^2} e^{-\frac{1}{2}\sqrt{\mathstrut 2\gamma}Q_0 r_H 
  \left( \sqrt{\mathstrut 2\gamma}Q_0 r_H +(-1)^b s_\varepsilon 2\mu\rho \right)}
  {\cal L}_{n_1,n_2} \left((-1)^b \sqrt{\mathstrut 2\gamma}Q_0 r_H +s_\varepsilon\mu\rho \right).
\end{eqnarray}
The last equality in eq.(\ref{eq:defJn1n2}) can be proved inductively. 
In this way, the quadratic GL free energy in the present case is expressed in the form
\begin{eqnarray} \label{eq:CPSF2}
  \frac{F_2}{2V} &=& \sum_{a,n_1,n_2} Y_{a,n_1}^\ast Y_{a,n_2} \biggl[ \left( \frac{(w^{-1})_{ss} 
  +(\tilde{w}^{-1})_{tt}}{2} -(-1)^a(\tilde{w}^{-1})_{st} \right) \delta_{n_1,n_2}
 \nonumber \\
  && \quad -2\int_{\rho_c}^\infty d\rho f(\rho) \Bigl\langle e^{-\frac{1}{2}|\mu|^2\rho^2}
  \Bigl\{ N_a (1+\varsigma)^2 {\rm Re}{\cal L}_{n_1,n_2}(\mu\rho) +N_b \varsigma^2 {\rm Re} 
  \left(e^{(-1)^{a+1}{\rm i}2 Q_0 v_x \rho} {\cal L}_{n_1,n_2}(\mu\rho) \right)
  \Bigr\} \Bigr\rangle \biggr]
 \nonumber \\
  && +\sum_{n_1,n_2} \biggl( Y_{1,n_1}^{\ast} Y_{2,n_2} \biggl[ -(N_1 +N_2) \, ({\tilde \delta} w)^{-1} \, 
  e^{-\gamma Q_0^2 r_H^2} {\cal L}_{n_2,n_1}(\ \sqrt[]{\mathstrut 2\gamma}Q_0 r_H)
 \nonumber \\
  && \quad +\int_{\rho_c}^\infty d\rho f(\rho) \Bigl\langle \varsigma(1+\varsigma) \Bigl\{ N_1 \left(
  J_{n_2,n_1}^\ast(\ \sqrt[]{\mathstrut 2\gamma}Q_0 r_H, \mu\rho)
  +J_{n_2,n_1}^\ast(\ \sqrt[]{\mathstrut 2\gamma}Q_0 r_H, -\mu\rho) \right)
 \nonumber \\
  && \quad \quad +N_2 \left( J_{n_1,n_2}(-\ \sqrt[]{\mathstrut 2\gamma}Q_0 r_H, \mu\rho)
  +J_{n_1,n_2}(-\ \sqrt[]{\mathstrut 2\gamma}Q_0 r_H, -\mu\rho) \right) \Bigr\} \Bigr\rangle \biggr] +({\rm c.c.}) \biggr), 
\end{eqnarray}
where 
\begin{equation}
  {\tilde \delta} w = - \frac{2(N_1 +N_2)}{(w^{-1})_{ss}-(\tilde{w}^{-1})_{tt}}
\end{equation}
is the measure of the $s$-$t$ mixing redefined within the treatment in this section. Here, since ${\tilde w}_{tt}$ is not the bare attractive interaction potential, the full mixing of the two pairing channels does not coincide with the limit in which ${\tilde \delta} w$ diverges. The above $F_2$-expression implies that the zero field transition temperature $T_c$ is determined from 
\begin{eqnarray}
  \frac{(w^{-1})_{ss}+(\tilde{w}^{-1})_{tt}}{2(N_1 +N_2)} &=& (1+\varpi^2)\int_{\rho_c}^\infty d\rho
  f(\rho)|_{T=T_c}) +\biggl[ \left(\varpi^2\int_{\rho_c}^\infty d\rho f(\rho)|_{T=T_c}-({\tilde \delta} w)^{-1} \right)^2
 \nonumber \\
  && \quad +\left(\frac{(\tilde{w}^{-1})_{st}}{N_1 +N_2} + \delta N\int_{\rho_c}^\infty d\rho f(\rho)|_{T=T_c} \right)^2
  \biggr]^{\frac{1}{2}},
\end{eqnarray}
where 
\begin{equation}
  \varpi=\frac{\tilde{J}}{4(1-\tilde{J})}. 
\end{equation}

In contrast to $F^{({\rm Q2D})}_2$, the paramagnetic effect in the present free energy does not disappear even when ${\tilde \delta} w$ diverges. Since the parameter $Q_0$ appears even in the diagonal terms with respect to ${\tilde \Delta}^{(0)}_a$, it is not easy to, in advance, prescribe the situation in which the orbital-limiting is realized. 

Again, the local approximation will be used for the quartic term of the corresponding GL free energy functional to determine stable vortex structures. The quartic GL term to be examined is 
\begin{equation} \label{eq:defCPSAbrikosov}
  \tilde{F}^{(R)}_4 \simeq \int d^3{\bf r} \sum_a N_a \bigl\langle |\Delta_a|^4 \bigr\rangle,
\end{equation}
where 
\begin{eqnarray} \label{eq:CPSGap4}
  |\Delta_a|^4 &=& (1+\varsigma)^4|\Delta^{(0)}_a|^4 +\varsigma^4|\Delta^{(0)}_b|^4
  -4\varsigma(1+\varsigma)^3|\Delta^{(0)}_a|^2{\rm Re}(\Delta^{(0)\ast}_a\Delta^{(0)}_b)
  -4\varsigma^3(1+\varsigma)|\Delta^{(0)}_b|^2{\rm Re}(\Delta^{(0)\ast}_b\Delta^{(0)}_a)
 \nonumber \\
  && +2\varsigma^2(1+\varsigma)^2\left[|\Delta^{(0)}_a|^2|\Delta^{(0)}_b|^2
  +{\rm Re}\left((\Delta^{(0)\ast}_b \Delta^{(0)}_a)^2\right)+|\Delta^{(0)\ast}_a \Delta^{(0)}_b|^2\right].
\end{eqnarray}

To examine eq.(\ref{eq:CPSGap4}) in more details, we first rewrite $\Delta^{(0)\ast}_a \Delta^{(0)}_{a^\prime} \ (a,a^\prime =1,2)$ in the form 
\begin{eqnarray} \label{eq:CPSGapFT}
  \tilde{\Delta}^{(0)}_{aa^\prime}({\bf G}) &=& \int d^2 {\bf r} \Delta^{(0)\ast}_a \Delta^{(0)}_{a^\prime}
  e^{-{\rm i}{\bf G}\cdot{\tilde {\bf r}}}
 \nonumber \\
  &=& \sum_{n_1,n_2} Y^{\ast}_{a,n_1} Y_{a^\prime,n_2}\int d^2 {\bf r}\varphi^\ast _{n_1}({\bf r}_a |0)
  \varphi_{n_2}({\bf r}_{a^\prime} |0) e^{-{\rm i}{\bf G}\cdot{\tilde {\bf r}}}
 \nonumber \\
  &=& (-1)^{m_1 m_2} \exp\biggl(-\frac{1}{2}|\Gamma_{aa^\prime}|^2 + (-1)^a {\rm i}Q_0 r_H^2 G_z \delta_{a,a^\prime} \biggr)
  \sum_{n_1,n_2}Y^{\ast}_{a,n_1}Y_{a^\prime,n_2} {\cal L}_{n_1,n_2}(\Gamma_{aa^\prime}),
\end{eqnarray}
where eq.(\ref{eq:fomBilinear}) was used, and
\begin{equation} \label{eq:defGammaaa}
  \Gamma_{aa^\prime}=-\frac{r_H}{\sqrt[]{\mathstrut 2}}\left( \left[G_x-(\delta_{a,a^\prime}-1)^a 2Q_0 \gamma^{1/2} r_H \right]
  +{\rm i} G_z \right).
\end{equation}
Using eqs.(\ref{eq:CPSGap4}) and (\ref{eq:CPSGapFT}), eq.(\ref{eq:defCPSAbrikosov}) becomes
\begin{eqnarray} \label{eq:CPSF4}
  \frac{2\tilde{F}^{(R)}_4}{N_1 +N_2} &=& \sum_{m_1 ,m_2} \left[ (1-\delta N)(1+3\varpi^2)
  +\frac{3}{4}\varpi^4 \right] |\tilde{\Delta}^{(0)}_{11}|^2
  +\left[ (1+\delta N)(1+3\varpi^2)+\frac{3}{4}\varpi^4 \right] |\tilde{\Delta}^{(0)}_{22}|^2
 \nonumber \\
  && \quad \quad -3\varpi^2 \left( \left[ 2(1-\delta N)+\varpi^4 \right]
  {\rm Re}(\tilde{\Delta}^{(0)}_{11}\tilde{\Delta}^{(0)}_{12})
  +\left[ 2(1+\delta N)+\varpi^4 \right] {\rm Re}(\tilde{\Delta}^{(0)}_{22}\tilde{\Delta}^{(0)}_{21}) \right)
 \nonumber \\
  && \quad \quad +\varpi^2 \left( 2+\frac{3}{2}\varpi^2 \right) \left[ \tilde{\Delta}^{(0)}_{11}\tilde{\Delta}^{(0)}_{22}
  +{\rm Re}(\tilde{\Delta}^{(0)\ast}_{12}\tilde{\Delta}^{(0)}_{21}) +|\tilde{\Delta}^{(0)}_{12}|^2 \right].
\end{eqnarray}
Substituting $Y_{a,n}$ determined from $F_2$ into eq.(\ref{eq:CPSGapFT}), the stable lattice structure can be determined from ${\tilde F}^{(R)}_4$. Further, just like in Figs.7 and 8 in Q2D case, we will present not only the resulting phase diagram but also the spatial variations of $|\Delta_a|$, $|\Delta_s|$, and $|\Delta_t|$ at some selected points in an $h$-$t$ phase diagram. According to the expressions given so far, they are given by 
\begin{eqnarray} \label{eq:AbsCPSGapa}
  && |\Delta_a|^2 =\left(1+\frac{\varpi^2}{2}\right)|\Delta^{(0)}_a|^2 +\frac{\varpi^2}{2}|\Delta^{(0)}_b|^2
  -\varpi^2{\rm Re}(\Delta^{(0)\ast}_1\Delta^{(0)}_2)
 \\ \label{eq:AbsCPSGaps}
  && |\Delta_s|^2 =\frac{1}{2} \left( |\Delta^{(0)}_1|^2 +|\Delta^{(0)}_2|^2
  +2{\rm Re}(\Delta^{(0)\ast}_1 \Delta^{(0)}_2) \right)
 \\ \label{eq:AbsCPSGapt}
  && |\Delta_t|^2 =\frac{1}{2(1-\tilde{J})} \left( |\Delta^{(0)}_1|^2 +|\Delta^{(0)}_2|^2
  -2{\rm Re}(\Delta^{(0)\ast}_1 \Delta^{(0)}_2) \right).
\end{eqnarray}

\begin{figure}[t]
\caption{(Color online) The resulting $h$-$t$ phase diagrams in the $s$-wave and $p$-wave mixed case for (a) $\delta N=0$ and (b) $\delta N=-0.1$ obtained in terms of eq.(\ref{eq:CPSGap}). The dashed portion of FOST2 is not identified due to a numerical difficulty. The used parameters are ${\tilde J}=0.35$ and $\delta w=5$.} \label{fig.9}
\end{figure}

\begin{figure}[b]
\caption{(Color online) Extended views of $|\Delta_2(z,x)|$ on FS2 at (A) (top) and (C) (bottom) in Fig.9(b).} \label{fig.10}
\end{figure}

\begin{figure}[t]
\caption{(Color online) The resulting $h$-$t$ phase diagram in the $s$-wave and $p$-wave mixed case obtained in terms of eq.(\ref{eq:CPSGap}). In the intermediate region between FOST2 and FOST6, the vortex lattice has a highly compressed triangular structure. The used parameters are ${\tilde J}=0.35$, $\delta N=0$, and $\delta w=0.454$.} \label{fig.11}
\end{figure}

A typical example of $\delta N$-dependences of the $h$-$t$ phase diagram at the same ${\tilde \delta}_R$-value is shown in Fig.9. The value ${\tilde \delta} w=5$ is estimated by assuming $\omega_c/T_c=10$ to roughly correspond to $\delta w=-2$, where $\omega_c$ is the higher energy cutoff for the pairing. For this reason, the $H_{c2}$-value suggested in Fig.9(a) is comparable with those in Fig.2 and is much smaller than the corresponding one in Q2D case with $\delta w=5$. In spite of this, the $\delta N$ dependence in Fig.9 with a more 3D-like FS is dramatic compared with that in Q2D case: Even a small $|\delta N|$ leads to an $H_{c2}(T)$ curve close to the orbital limit, although we have checked that the $H_{c2}(T)$ curve in Fig.9(b) lies slightly below that in the orbital-limited case. 

More important differences from those in Q2D case are seen in the resulting vortex lattice structures. According to the results in Q2D case, a slight inclusion of a finite $\delta w$ has similar roles to those of a finite $\delta N$, and the scenario suggested by Fig.3 was that the states of LO type with no reflection symmetry are destabilized. However, the corrugation of the cylindrical FS, or a 3D-like FS seems to destabilize rather the novel striped modulation induced by the finite $\delta N$ and appeared as (B) and (C) in Fig.3(b). In fact, it seems based on some phase diagrams we have numerically obtained that the structures (B) and (C) in Fig.3(b) with reflection symmetry are close in energy to another structures in intermediate fields, (B) and (C) in Fig.3(a) with no reflection symmetry and thus that the crossover between the former structures competes with the corresponding one between the latter structures. The former is supported in part by the nearly straight cylindrical FS, while the corrugation of the cylindrical FS or a 3D-like FS favors the latter. We stress that such a competition is absent in the single pairing case in sec.III because it is the momentum dependence in eq.(\ref{eq:deltaa}) which induces such a competition between two kinds of modulated states. 

In fact, Fig.9(a) should be compared with Fig.7: For instance, the square lattice near $H_{c2}(0)$ in Fig.7 is replaced in Fig.9(a) by that of LO type, and the roles of the two kinds of structures (one with reflection symmetry and the other of LO type with no reflection symmetry) in intermediate fields in Fig.7 are precisely exchanged in Fig.9(a). Further, Fig.9(b) shows that, with increasing $|\delta N|$, the region of the triangular lattice in intermediate fields shrinks in contrast to the strongly anisotropic triangular lattice near $T_c$, and that the high field region above FOST1 in Fig.9(a) disappears. Consequently, far from $T_c$, the only stable structure in Fig.9(b) is a strongy anisotropic and {\it tilted} vortex lattice with no reflection symmetry. As Fig.10 show, the vortices in Fig.9(b) have cores compressed along the $c$-axis. With decreasing the field, the tilt angle of the stripes in Fig.9(b), which is a vestige of the FFLO modulation in (A) of Fig.9(a), decreases. Since the lattice structure there is close to the square symmetry rather than the hexagonal one, however, an FOST, just like FOST2 in Fig.3(a), inevitably occurs to transform into the triangular lattice (C). This example also indicates a similar role of ${\tilde \delta} w$ to that of $\delta N$. We also note that phase diagrams similar to Fig.9(b) have been obtained quite often in our numerical calculations. For instance, even for ${\tilde \delta} w = 0.4$ which seems to be a value closer to the orbital limiting, we have obtained the results similar to Fig.9(b) irrespective of the used $\delta N$-value. 

We have not examined a phase diagram for quite a small $|\delta N|$ interpolating Fig.9 (a) and (b). Based on the above-mentioned similar roles of ${\tilde \delta} w$ and $\delta N$, however, it is valuable to examine the $\delta N=0$ case with a larger $s$-$t$ mixing than that in Fig.9(a). For this reason, we show such an example in Fig.11. It seems, except the presence of FOST4 and FOST5 there, that Fig.11 interpolates between Fig.9(a) and (b). In fact, these two FOSTs are expected to change into crossovers once $\delta N$ becomes nonzero, because the roles of even and odd LLs are exchanged through FOST4 and FOST5, and an even-odd LL mixing due to a nonzero $\delta N$, as in Fig.3, tends to change an FOST into a crossover. Therefore, it is natural to expect that, in the situations interpolating Fig.9(a) and (b), the {\it intermediate} triangular lattices are simply lost with increasing $|\delta N|$. 

\section{V Summary and Discussion}

In this paper, possible vortex lattice structures in noncentrosymmetric Rashba superconductors have been studied, and, as a result of the anisotropic Zeeman effect peculiar to Rashba superconductors, the vortex structure was found to change depending on the pairing symmetry. 
Through our calculations for several model pairing states, three types of sequences of field-induced structural crossovers have been found to appear in superconductors with the ASOC of pure Rashba type in intermediate fields depending on the value of a normalized ASOC and on the pairing state: 1) LO-like structure with {\it no} reflection symmetry and a field-induced rotation of its orientation, 2) novel striped lattice modulating along the ${\bf Q}_0$-direction and its crossover to a compressed square lattice, and 3) intermediate triangular lattices {\it differing} from the familiar one near $T_c$ and in lower fields. However, the LO structure occurring in the $d_{x^2-y^2}$-pairing case under a field parallel to the gap 
nodes is exceptional and has a modulation parallel to the $c$-axis and reflection symmetry. 

To obtain close correlations between the vortex structure and pairing symmetry, a detailed analysis taking account of a more realistic band structure will be neccesary. Nevertheless, based on the numerical study we have performed so far, the following two conclusions are expected to be unaffected by refining the starting microscopic model. Below, we focus on realistic cases with nonvanishing $|\delta N| \sim 0.1$. 

In the case where the $s$-$t$ mixing is negligible, a modulated state with weak stripes parallel to the $c$-axis or an anisotropic square vortex lattice is realized in the intermediate and high field ranges, depending on the situation, and is expected to have reflection symmetry. In the presence of vertical line gap-nodes parallel to the $c$-axis, the resulting high field state is affected by the gap nodes and may be the LO-like vortex lattice with stripes perpendicular to the $c$-axis if the field is parallel to a gap node. 

In the case with a significant amount of $s$-$t$ mixing, the paramagnetic depairing effect is reduced irrespective of the pairing state, leading to an enhancement of $H_{c2}$, while the vortex lattice structure seems to depend on the dimensionality of the Fermi surface: For the 2D-like case in which FS takes the form of a nearly straight cylinder, the resulting vortex lattices seem to always have reflection symmetry and to yield the structural crossover 2) indicated above. In the case with a more realistic FS such as a corrugated cylinder, however, the resulting vortex lattice in higher fields has no reflection symmetry reflecting the crossover 1) suggesting the presence of the LO-like state in $\delta N \to 0$ limit. These scenarios are not satisfied in superconductors with the vertical line gap-nodes ($\parallel c$-axis) and under a field parallel to a gap-node, where the tilted structures with no reflection symmetry do not appear, reflecting a pinning of the striped structure via the gap nodes. 

Finally, comments relevant to real experiments are in order: 

Experimentally, an imaging of a vortex lattice can be seen, for instance, through neutron scattering measurements. In such an experiment, however, the structure is detected as a flux density distribution which, in turn, is determined by a spatial distribution of the supercurrent. Although we have not calculated the supercurrent density in the present work, the resulting flux density is, roughly speaking, proportional to the summation $\sum_{a=1,2} |\Delta_a|^2$ so that the spatial patterns shown in the figures in the preceding sections are essentially detectable. 

Among the existing noncentrosymmetric superconductors, CeRhSi$_3$ \cite{Kimura} and CeIrSi$_3$ \cite{Settai} seem to correspond to the case with a negligibly small $s$-$t$ mixing because they show significantly reduced $H_{c2}(T)$ in ${\bf H} \perp c$ compared with that in ${\bf H} \parallel c$. If, as in the pairing model proposed recently \cite{Tada}, the pairing state has no vertical line gap-nodes, the present results imply that the resulting vortex lattices should keep reflection symmetry, and that no unusual vortex dynamics is expected to appear in these materials (see below).  

By contrast, in CePt$_3$Si, nearly isotropic $H_{c2}$-curves \cite{jpsj131} have been previously determined experimentally which suggest that the paramagnetic depairing is weak in this material. Based on the present results, this implies that its pairing state has a significant $s$-$t$ mixing, or that the bare paramagnetic effect is negligibly weak. If the latter possibility is correct, a natural guess is that the vortex state will be a {\it conventional} orbital-limited one, which cannot explain the recent interesting observation of an extremely small magnetic decay rate in CePt$_3$Si in ${\bf H} \parallel a$  \cite{Mota} without extrinsically assuming the presence of twin boundaries. On the contrary, if the former case is valid, and the pairing state has no vertical line gap-nodes, the resulting vortex lattice has no reflection symmetry (see Fig.9(b)). This suggests the presence of two domains of vortex lattices in real CePt$_3$Si \cite{comAF}, although the pairing state of this material is expected to have time reversal symmetry, and it is possible that the observation in Ref.\cite{Mota} is intrinsically explained without invoking \cite{Sigrist} extrinsic twin boundaries . To clarify this point, similar magnetic measurements in ${\bf H} \parallel c$ and in other Rashba superconductors such as CeRhSi$_3$ and CeIrSi$_3$ are to be performed.

\end{document}